\documentclass[journal]{IEEEtran}

\ifCLASSINFOpdf
\else
\fi
\usepackage{graphicx}
\usepackage{cite}
\usepackage{picinpar}
\usepackage{amsmath}
\usepackage{url}
\usepackage[latin1]{inputenc}
\usepackage{colortbl}
\usepackage{soul}
\usepackage{multirow}
\usepackage{pifont}
\usepackage{color}
\usepackage{alltt}
\usepackage{enumerate}
\usepackage{siunitx}
\usepackage{epstopdf}
\usepackage{pbox}

\usepackage{amsfonts,amssymb}
\usepackage{diagbox}
\usepackage{subfigure}
\usepackage{booktabs}

\graphicspath{{./}}

\begin{document}

\title{A Feature Weighted Mixed Naive Bayes Model for Monitoring Anomalies in the Fan System of a Thermal Power Plant}

\author{Min~Wang,~\IEEEmembership{Graduate Student Member,~IEEE,}
        Li~Sheng,~\IEEEmembership{Member, IEEE,}
        Donghua~Zhou,~\IEEEmembership{Fellow,~IEEE,}
        and~Maoyin~Chen,~\IEEEmembership{Member, IEEE}
\thanks{This work was supported by the National Natural Science Foundation of China under Grant 62033008, 61873143. (Corresponding
author: Donghua Zhou, Maoyin Chen.)}

\thanks{Min~Wang, and Maoyin~Chen are with the Department of Automation, Tsinghua University, Beijing 100084, China (e-mail:m-wang18@mails.tsinghua.edu.cn, mychen@tsinghua.edu.cn).}
\thanks{Li Sheng is with the College of Control Science and Engineering, China University of Petroleum (East China), Qingdao 266580, China (e-mail:shengli@upc.edu.cn).}
\thanks{Donghua Zhou is with the College of Electrical Engineering and Automation, Shandong University of Science and Technology, Qingdao, Shandong 266590, China, and also with the Department of Automation, Tsinghua University, Beijing 100084, China (e-mail:zdh@tsinghua.edu.cn).}
}

\markboth{IEEE/CAA JOURNAL OF AUTOMATICA SINICA}%
{Shell \MakeLowercase{\textit{et al.}}: Bare Demo of IEEEtran.cls for IEEE Journals}

\maketitle

\begin{abstract}
With the increasing intelligence and integration, a great number of two-valued variables (generally stored in the form of 0 or 1) often exist in large-scale industrial processes. However, these variables cannot be effectively handled by traditional monitoring methods such as linear discriminant analysis (LDA), principal component analysis (PCA) and partial least square (PLS) analysis. Recently, a mixed hidden naive Bayesian model (MHNBM) is developed for the first time to utilize both two-valued and continuous variables for abnormality monitoring. Although the MHNBM is effective, it still has some shortcomings that need to be improved. For the MHNBM, the variables with greater correlation to other variables have greater weights, which can not guarantee greater weights are assigned to the more discriminating variables. In addition, the conditional probability ${P\left( {{{x}_{j}}\left| {{{x}_{j'}},{y} = k} \right.} \right)}$ must be computed based on historical data. When the training data is scarce, the conditional probability between continuous variables tends to be uniformly distributed, which affects the performance of MHNBM. Here a novel feature weighted mixed naive Bayes model (FWMNBM) is developed to overcome the above shortcomings. For the FWMNBM, the variables that are more correlated to the class have greater weights, which makes the more discriminating variables contribute more to the model. At the same time, FWMNBM does not have to calculate the conditional probability between variables, thus it is less restricted by the number of training data samples. Compared with the MHNBM, the FWMNBM has better performance, and its effectiveness is validated through numerical cases of a simulation example and a practical case of the Zhoushan thermal power plant (ZTPP), China.

\end{abstract}


\IEEEpeerreviewmaketitle

\section{Introduction}
\label{Introduction}
\IEEEPARstart{W}{ith} increasing intelligence and integration, a great number of two-valued variables (generally stored as 0 or 1 value) often exist in large-scale industrial processes. For instance, $17381$ variables are monitored in the No. 1 generator unit of the Zhoushan thermal power plant (ZTPP), where two-valued variables are more than $8820$. These two-valued variables mainly include status monitoring variables and numerical range variables, such as control command signals and vibration over-limit signals, which switch from one state to the other with less influence from process fluctuation noise.

In order to insure the high safety and reliability of large-scale industrial processes, the problem of monitoring anomalies becomes more and more important\cite{2019AYang,2020LatentZhu,2018Unsupervised,2020Dynamic,2017Incipient}. The timely and accurate abnormal monitoring can effectively reduce waste of resources, economic losses, and even casualties \cite{Kai2020Data,Anagnostou2018Observer,Zhang8859224,Ji2019Incipient,2020Distributed,2020Weighted}. Among a large number of monitoring methods, data-driven techniques have attracted much attention with the advantages of requiring less system information and prior knowledge than model-based and expert experience methods \cite{2018AnZhang,Yin7509639,2013Review,Yao2019Scalable,Agrawal2015Review,Gao2015A,Huang2019Nonlinear,MinWang2020IEEE}. For example, principal component analysis (PCA) and its variants have been widely used in industrial processes \cite{Ricardo1996Identification,Li2000Recursive}.  In order to detect quality-related faults, approaches based on partial least square (PLS) analysis have been proposed \cite{Qin1998Recursive,2016Comprehensive}. When the training data contains both normal and abnormal working condition samples, linear discriminant analysis (LDA) has been utilized \cite{He2005A}. In addition, many other machine learning methods, such as K-nearest neighbors (KNN) \cite{He2007Fault}, support vector machine (SVM) \cite{Achmad2007Support}, etc., have also been applied in abnormal monitoring.

However, the fact that two-valued variables ubiquitously exist in large-scale industrial processes presents a challenge to traditional monitoring methods. It is well known that the above mentioned methods are strongly based on continuous variables and may be not suitable for two-value variables. For example, PCA, PLS, LDA, \textit{etc.} obtain a subspace that is convenient for monitoring through decomposition and then construct statistics or hyperplanes. But these operations are based on Euclidean distance or Mahalanobis distance, which can't effectively mine the process information of two-valued variables. Two-valued variables are usually deleted during the data preprocessing stage \cite{GeZhiqiangIEEEAccess,MinWang2020CEP}. Recently, the mixed hidden naive Bayesian model (MHNBM) was proposed for the first time to combine both two-valued and continuous variables to improve monitoring performance \cite{MinWang2020CEP}. Although MHNBM is effective, the variables with greater correlation to other variables have greater weights, which can not guarantee that greater weights are assigned to the more discriminating variables. Moreover, the conditional probability ${P\left( {{{x}_{j}}\left| {{{x}_{j'}},{y} = k} \right.} \right)}$ between ${x}_{j}$ and ${ x}_{j'}$ under ${ y} = k$ must be computed based on the historical data. When training data is scarce, the conditional probability between continuous variables tends to be uniformly distributed, which will affect performance.

Motivated by the above discussions, a model known as the feature weighted mixed naive Bayes model (FWMNBM) is proposed to overcome the shortcomings of MHNBM. In FWMNBM, the variables that are more correlated to the class have greater weights which results in variables with greater differences under different working conditions contribute more to the model. Meanwhile, FWMNBM can avoid calculating the conditional probability between variables such that it can still be used when there is not enough training data. In addition, a more effective consistent characterization technique is developed for the correlation of mixed variables, and the corresponding feasibility analysis is conducted. Compared with MHNBM, FWMNBM has better performance, and the effectiveness of FWMNBM is validated through the simulations of a numerical example and a practical vibration fault case.

In this paper, the remainder is organized as follows. Some preliminaries are briefly outlined in Section \ref{Preliminary}. In Section \ref{Mixedhiddennaivebayesianmodel}, FWMNBM is elaborated on. The estimation of parameters is introduced in Section \ref{ParametersEstimation}. In Section \ref{Casestudies}, the effectiveness of FWMNBM is verified. Finally, conclusions are drawn in Section \ref{Conclusion}.

\section{Preliminary}
\label{Preliminary}

In this section, the preliminary for FWMNBM is introduced. Let ${\pmb X} = {( \pmb x_i)_{1 \le i \le n}} \in {\mathbb{R}^{n \times p  }}$ be the training data set which contains $n$ instances. Here ${\pmb x_i}{\rm{ = }}\left( {{x_{i1}},{x_{i2}}, \ldots ,{x_{ip}}} \right) \in {\mathbb{R}^p}$ is the sampled data at time $i$. From the assumption of independent and identical distribution, it can be determined that \cite{Ozuysal2010Fast}:
\begin{align}\label{PreliminaryConditionalProbability}
 P( {{{\pmb X}} | {y = k} } ) = \mathop \Pi \limits_{j = 1}^p P( {{x_{j}} | {{y} = k} }),
\end{align}
where  $x_j$ is the $j$th variable of $\pmb X$, $y$ denotes the label variable and ${y_i} \in \left\{ {1, \ldots ,k,\ldots ,K} \right\}$ is the class of instance ${\pmb x_i}$.

For a continuous variable, suppose the variable obeys Gaussian distribution defined as
\begin{align}\label{PreliminarycontinuousConditionalProbability}
P ( {{x_j} | {{y} = k} } ) = \frac{1}{{\sqrt {2\pi \sigma _{kj}^2} }}\exp ( { - \frac{{{{ ( {{x_j} - {\mu _{kj}}} )}^2}}}{{2\sigma _{kj}^2}}} ),
\end{align}
where ${\mu _{kj}}$ is the mean of the $j$th variable under the $k$class, and $\sigma _{kj}^2$ is the corresponding variance. If $x_j$ is a two-valued variable, the Bernoulli distribution is introduced as follows \cite{Fortuny2018Wallenius}:
\begin{align}\label{PreliminarybinaryConditionalProbability}
{P} ( {{x_{j}} | {{ y} = k} } ) = \theta _{kj}^{{x_{j}}}{ ( {1 - {\theta _{kj}}}  )^{1 - {x_{j}}}},
\end{align}
where ${\theta _{kj}}{\rm{ = }}P\left( {{x_j} = 1\left| {{y} = k} \right.} \right)$ is the response probability.

When $\pmb X $ only contains two-valued variables or continuous variables, the label of a new sampled data $\pmb x_{new}$ is the maximum posterior probability denoted by
\begin{align}\label{PreliminaryNBposteriorprobability}
P ( {{y} = k | \pmb X \leftarrow {\pmb x_{new}}  } ) = \frac{{P ( {\pmb X | {{ y} = k}  } )P ( {{y} = k} )}}{\sum\limits_{k' = 1}^K {P ( {\pmb X | {{y} = k'}  }  )P ( {{ y} = k'} )} },
\end{align}
where ${p_k}{\rm{ = }}P ( {{ y} = k} )$ is the prior probability, $\pmb X \leftarrow {\pmb x_{new}}$ indicates that $\pmb X$ is replaced by $\pmb x_{new}$.

\section{Main Algorithm}
\label{MainAlgorithm}
\subsection{FWMNBM}
\label{Mixedhiddennaivebayesianmodel}

If two-valued variables and continuous variables exist in $\pmb X $ at the same time, let ${{\pmb X}} = [ {{{ {{{\pmb X}_t}} }},{{{{{\pmb X}_c}}}}} ]$, where ${{\pmb X}_c}  \in {\mathbb{R}^{{n \times p_1}}}$ is the continuous data set, ${{\pmb X}_t} \in {\mathbb{R}^{{n \times p_2}}}$ is the two-valued variable set, where $ {p = {p_1} + {p_2}} $. Under the assumption that the variables are independent, and assume that the continuous variable (${ x_j} \in {{\pmb X}_c}$) obeys the Gaussian distribution as equation (\ref{PreliminarycontinuousConditionalProbability}) which is denoted by ${P_c} ( {{x_j} | {{y} = k}  }  )$, and two-valued variable (${ x_j} \in {{\pmb X}_t}$) obeys the Bernoulli distribution as equation (\ref{PreliminarybinaryConditionalProbability}) which is denoted as ${P_t} ( {{x_j} | {{y} = k} } )$. Then equation (\ref{PreliminaryConditionalProbability}) can be written into
\begin{align}\label{mixedConditionalProbability}
P ( {{{\pmb X}} | {{ y} = k} } )
	= \mathop \Pi \limits_{j = 1}^{{p_1}} {P_c}\left( {{x_j}\left| {{y} = k} \right.} \right)\mathop \Pi \limits_{j = 1}^{{p_2}} {P_t}\left( {{x_j}\left| {{y} = k} \right.} \right).
\end{align}

After all parameters ${\mu _{kj}}$, $\sigma _{kj}^2$, ${\theta _{kj}}$, ${p_k}$ are estimated respectively based on the historical data, the posterior probability (\ref{PreliminaryNBposteriorprobability}) of a new sample $\pmb x_{new}$ is
\begin{align}\label{NBposteriorprobability} 
&P( {y = k| \pmb X \leftarrow {{\pmb x_{new}}}}) \notag \\
&= \frac{{P ( {y = k} )\prod\limits_{j = 1}^{{p_1}} {{P_c} ( {{ x_j} | {y = k} } )\prod\limits_{j = 1}^{{p_2}} {{P_t} ( {{x_j} | {y = k} } )} } }}{{\sum\limits_{k' = 1}^K {P ( { y = k'} )\prod\limits_{j = 1}^{{p_1}} {{P_c} ( {{x_j} | {y = k'} } )\prod\limits_{j = 1}^{{p_2}} {{P_t} ( {{x_j} | {y = k'} } )} } } }} .
\end{align}

In the actual process, the independence assumption of variables is too stringent. Among numerous approaches alleviating this assumption, feature weighting is a better choice. Inspired by \cite{Zhang2016Two}, the model (\ref{mixedConditionalProbability}) could be amended as follows:
\begin{align}\label{HMNBMprobability}
P ( {{{\pmb X}}  | {{y} = k} } ) &= P' ( {{{\pmb X}}  | {{y} = k} }  ) =\mathop \Pi \limits_{j = 1}^p P' ( {{x_j} | {{y} = k} } )  \notag \\
 &= \mathop \Pi \limits_{j = 1}^{{p_1}} {P'_c} ( {{x_j} | {{y} = k} }  )\mathop \Pi \limits_{j = 1}^{{p_2}} {P'_t} ( {{x_j} | {{y} = k} } ),
\end{align}
where
\begin{align}\label{weightpostprobability1}
P' ( {{x_j} | {{y} = k} } ) =  {P ( {{x_j} | {{y} = k} } )}^{FW_{j}},
\end{align}
where $FW_{j}$ is the feature weight of the $j$th feature. It should be noted that variables more correlated with the class should have greater weights, and uncorrelated attributes should have smaller contributions. Therefore, $FW_{j}$ should reflect the correlation between the $x_j$ and $y$ as accurately as possible. The mutual information (MI) ${MI ( {{x_j},{y}} )}$ is used to characterize the correlation between $x_j$ and $y$.  ${MI ( {{x_j},{y}} )}$ can effectively describe the correlation between $x_j$ and $y$, but it also contains some correlational information between $x_j$ and other variables (such as $x_{j'}$) because variables are coupled. Then the average feature-feature intercorrelation is introduced to compute the feature weight \cite{8359364} 
\begin{align}\label{correlationindex}
C{I_j} = {MI ( {{x_j}, y} )} - {\frac{1}{p-1}} \sum\limits_{j = 1,j \ne j'}^p {MI ( {{x_j},{x_{j'}}} )},
\end{align}
where ${MI\left( {{x_j},{x_{j'}}} \right)}$ is MI between the $j$th and $j'$th variables. The value of $C{I_j}$ may be negative, but the weight must be non-negative. Thus, the final form of the weight of feature ($F{W_{j}}$) is transformed as
\begin{align}\label{transformefunction}
F{W_j} = {1 \mathord{\left/
 {\vphantom {1 { ( {1 + {e^{C{I_j}}}} )}}} \right.
 \kern-\nulldelimiterspace} { ( {1 + {e^{-C{I_j}}}} )}}.
\end{align}

The feature weights are normalized to satisfy
\begin{align}\label{normalizedfeatureweights}
\sum\limits_{j = 1}^p FW_{j} = 1.
\end{align}

In order to effectively describe the correlation between two-valued and continuous variables, continuous variables are processed as follows.

\textbf{Definition 1.} If ${x_j} \in {\pmb {X}_c}{\kern 1pt} {\kern 1pt}$ , let ${{{ x}'}_j}:\left\{ {{{x'}_{1j}},} \right.{{x'}_{2j}}, \ldots ,$ $\left. {{{x'}_{nj}}} \right\}$ be the auxiliary two-valued variable corresponding to ${x_j}:\left\{ {{x_{1j}},{x_{2j}}, \ldots ,{x_{nj}}} \right\} $ , which can be obtained through clipping \cite{Ratanamahatana2005A} as  
\begin{align}\label{clippingresult}
{x'_{ij}} = \left\{ {\begin{array}{*{20}{c}}
{1,{\kern 1pt} {\kern 1pt} {\kern 1pt} {\kern 1pt} {\kern 1pt} {\kern 1pt} {x_{ij}} > {\kern 1pt} {{\sum\limits_{k = 1}^K { ( {{u_{kj}}{n_k}} )} } / n}},\\
{0,{\kern 1pt} {\kern 1pt} {\kern 1pt} {\kern 1pt} {\kern 1pt} {\kern 1pt} {x_{ij}} \le {\kern 1pt} {{\sum\limits_{k = 1}^K { ( {{u_{kj}}{n_k}} )} } / n}},
\end{array}} \right.
\end{align}
where ${n_{k}} = \sum\limits_{i = 1}^n {{\pi _{ik}}}$. Then ${x'_j}$, instead of ${ x_j}$, is used to build the correlation index. The feasibility analysis is given in Appendix \ref{AppendixAProofofDefinition1}.

The MI between $x_j$ and $x_{j'}$ can be computed by \cite{Yuanfang1991Conditional}:
\begin{align}\label{mutualinformation}
MI ( {{x_j},{x_{j'}}} ) =  \sum\limits_{{x_j},{x_{j'}}} {P ( {{x_j},{x_{j'}}}  )\log \frac{{P ( {{x_j},{x_{j'}}} )}}{{P ( {{x_j}} )P ( {{x_{j'}}} )}}}.
\end{align}

Then equation (\ref{NBposteriorprobability}) can be transformed to
\begin{align}\label{FeatureWeightedmixedposteriorprobability}
&P ( {y = k | \pmb X \leftarrow {{\pmb x_{new}}} } )= [P ( y = k ) \prod\limits_{j = 1}^{p_1} {P_c}{{ ( {{ x_j} | {y = k} } )}^{F{W_j}}} \notag \\
&\times \prod\limits_{j = 1}^{{p_2}} {{P_t}{{ ( {{x_j} | {y = k} }  )}^{F{W_j}}}}] [\sum\limits_{k' = 1}^K ( P ( { {y = k'} } ) \notag \\
&\times  \prod\limits_{j = 1}^{{p_1}} {{P_c}{{ ( {{x_j} | {y = k'} }  )}^{F{W_j}}}\prod\limits_{j = 1}^{{p_2}} {{P_t}{{ ( {{x_j} | {y = k'} } )}^{F{W_j}}}} })]^{-1}.
\end{align}

$\sum\limits_{k' = 1}^K  P ( { {y = k'} } ) \prod\limits_{j = 1}^{{p_1}} {{P_c}{{ ( {{x_j} | {y = k'} }  )}^{F{W_j}}}\prod\limits_{j = 1}^{{p_2}} {{P_t}{{ ( {{x_j} | {y = k'} } )}^{F{W_j}}}} }$ is constant for each $k$, then the label of new instance (${\pmb x_{new}}$ ) is
\begin{align}\label{newlysampleddatalabel1}
	{y_{new}} &= \mathop {\arg \max }\limits_k {\kern 1pt} {\kern 1pt} {\kern 1pt} {\kern 1pt} {\kern 1pt} {\kern 1pt} P ( { y = k | \pmb X \leftarrow {{\pmb x_{new}}} } ) \notag \\
	&= \mathop {\arg \max }\limits_k [P ( {y = k} )\prod\limits_{j = 1}^{{p_1}} {{P_c}{{ ( {{x_j} | {y = k} } )}^{F{W_j}}}} \notag \\
	&~~~~~~~~~~~~~~~\times \prod\limits_{j = 1}^{{p_2}} {{P_t}{{ ( {{x_j} | {y = k}  } )}^{F{W_j}}}} ] \notag \\
	&= \mathop {\arg \max }\limits_k \{ \ln P ( {y = k} ) \notag \\
	&~~~~~~~~~~~~~~~ + \sum\limits_{j = 1}^{{p_1}} {F{W_j}} \ln {P_c} ( {{x_j} | {y = k} } )\notag \\
	&~~~~~~~~~~~~~~~ + \sum\limits_{j = 1}^{{p_2}} {F{W_j}} \ln {P_t} ( {{x_j} | {y = k}} )  \}.
\end{align}

Denote the estimations of the parameters $\mu_{kj},$ $\sigma_{kj}^2,$ $\theta_{kj},$ $p_{k}$ as $\hat{\mu}_{kj},$ $\hat{\sigma}_{kj}^2,$ $\hat{\theta}_{kj},$ $\hat{p}_{k},$ respectively. Suppose that these parameters are estimated based on the historical data and feature weights are calculated. Then we have
\begin{align}\label{newlysampleddatalabel21}
&{y_{new}} = \mathop {\arg \max }\limits_k \{ {\ln {{\hat p}_k} + }\sum\limits_{j = 1}^{{p_2}} F{W_j} [ {{ { ( {{\pmb x_{new}}} )_j} }\ln {{\hat \theta }_{kj}}} \notag \\
&~~+  ( {1 - { { ( {{\pmb x_{new}}} )_j} }} )\ln  ( {1 - {{\hat \theta }_{kj}}}  )]  \notag \\
&~~+ {\sum\limits_{j = 1}^{{p_1}} {F{W_j}\ln  [ {\frac{1}{{\sqrt {2\pi \hat \sigma _{kj}^2} }}\exp  ( { - \frac{{( {{ { ( {{\pmb x_{new}}} )_j} } - {{\hat \mu }_{kj}}} )^2}}{{2\hat \sigma _{kj}^2}}}  )} ]} } \},
\end{align}
where ${\left( {{\pmb x_{new}}} \right)_j}$ is the $j$th feature of $\pmb x_{new}$.

Denoting that
\begin{align}\label{vecterx}
\tilde {\pmb x} = [ {{ ( {{\pmb x_{new}}} )_1},{ ( {{\pmb x_{new}}} )_2}, \ldots ,{ ( {{\pmb x_{new}}} )_{{p_2}}},1} ],
\end{align}
\begin{align}\label{vecterthita}
&{{\pmb \varphi} _k} = [ F{W_1} \ln \frac{{{{\hat \theta }_{k1}}}}{{1 - {{\hat \theta }_{k1}}}}, \ldots , F{W_{p_2}} \ln \frac{{{{\hat \theta }_{k{p_2}}}}}{{1 - {{\hat \theta }_{k{p_2}}}}}, \notag \\
&~~~~~~~~~~~~~~~~~~\ln {{\hat p}_k} + \sum\limits_{j = 1}^{{p_2}} F{W_j} {\ln ( {1 - {{\hat \theta }_{kj}}} )}]^T,
\end{align}
\begin{align}\label{vectermeanvar}
{\phi _k} = \sum\limits_{j = 1}^{{p_1}} F{W_j} {[ { - \frac{1}{2}\ln ( {2\pi \hat \sigma _{kj}^2} ) - \frac{{( {{{ ( {{\pmb x_{new}}} )_j}} - {{\hat \mu }_{kj}}})^2}}{{2\hat \sigma _{kj}^2}}} ]},
\end{align}
we have
\begin{align}\label{newlysampleddatalabel2}
{y_{new}} &= \mathop {\arg \max }\limits_k \{ {\sum\limits_{j = 1}^{{p_2}} {{{ ({{\pmb x_{new}}} )_j}}F{W_j}\ln \frac{{{{\hat \theta }_{kj}}}}{{1 - {{\hat \theta }_{kj}}}}} }  \\
&~~~~+ \ln {{\hat p}_k} + \sum\limits_{j = 1}^{{p_2}} {F{W_j}\ln ( {1 - {{\hat \theta }_{kj}}} )} \notag \\
&~~~~+ {\sum\limits_{j = 1}^{{p_1}} {F{W_j} ( { - \frac{1}{2}\ln ( {2\pi \hat \sigma _{kj}^2} ) - \frac{{ ( {{{ ({{\pmb x_{new}}} )_j}} - {{\hat \mu }_{kj}}}  )}^{2}}{{2\hat \sigma _{kj}^2}}} )} } \}  \notag \\
&= \mathop {\arg \max }\limits_k ( {\tilde {\pmb x} \cdot {{ \pmb \varphi} _k} + {\phi _k}} ).
\end{align}

\subsection{Parameters Estimation}
\label{ParametersEstimation}

In this subsection, $\pmb X$ is used for parameter estimation. According to maximum likelihood estimation (MLE) \cite{Collins2004Parameter}, the prior probability can be given as
\begin{align}\label{pkbymle}
{\hat p_k} = \hat p_k^{MLE} = {{\sum\limits_{i = 1}^n {{\pi _{ik}}} }/ n},
\end{align}

where ${\pi _{ik}} = 1 \{ {{y_i} = k} \}$.

For two-valued variables,  the response probability is represented by
\begin{align}\label{thetakjbymle}
{\hat \theta _{kj}} = \hat \theta _{kj}^{MLE} = ({\sum\limits_{i = 1}^n {{x_{ij}}{\pi _{ik}}} }) / \sum\limits_{i = 1}^n {{\pi _{ik}}},
\end{align}
where ${x_{ij}}$ is the $i$th sample of the $x_j$. In order to avoid a probability of 0 or 1, the estimated response probability and prior probability are truncated as following \cite{Balakrishnan2008Bernoulli}.

Assuming $\sum\limits_i^n {{\pi _{iK}}}  = \mathop {\max }\limits_{1 \le k \le K} \sum\limits_i^n {{\pi _{iK}}}$, then for $1 \le k \le K - 1$, we have
\begin{align}\label{pkbydte}
{\hat p_k} = \hat p_k^{DTE} = \max \{ {\xi ,\min  ( {{{\sum\limits_{i = 1}^n {{\pi _{ik}}} } / n}, 1 - \xi} )} \},
\end{align}
where $\xi $ is a small positive value ($\xi = 10^{-6}$ in this article). When $k = K$, we can get $\hat p_K$ by
\begin{align}\label{pcbydte}
{\hat p_K} = \hat p_K^{DTE} = 1 - \sum\limits_{k = 1}^{K - 1} {{{\hat p}_k}} .
\end{align}

Under the same assumption, we have ${\theta _{Kj}} = \mathop {\max }\limits_{1 \le k \le K} {\kern 1pt} {\kern 1pt} {({\sum\limits_{i = 1}^n {{x_{ij}}{\pi _{ik}}} }) / { {\sum\limits_{i = 1}^n {{\pi _{ik}}} } }}$, for $1 \le k \le K - 1$. Then the response probability is defined by
\begin{align}\label{thetakjbydte}
{\hat \theta _{kj}}{\rm{ = }}\hat \theta _{kj}^{DTE}{\rm{ = }}\max \{ {\xi ,\min [ {(\sum\limits_{i = 1}^n {{x_{ij}}{\pi _{ik}}}) /{{\sum\limits_{i = 1}^n {{\pi _{ik}}} }}, 1 - \xi } ]} \} .
\end{align}

When $k = K$, one has
\begin{align}\label{thetakcbydte}
{\hat \theta _{Kj}} = \hat \theta _{Kj}^{DTE} = 1 - \sum\limits_{k = 1}^{K - 1} {\hat \theta _{kj}^{DTE}},
\end{align}
where $\hat p_k^{DTE} \in [ {\xi ,1 - \xi } ],\hat \theta _{kj}^{DTE} \in  [ {\xi ,1 - \xi } ]$.

For continuous variables, the mean and the standard deviation (std) are respectively  estimated as
\begin{align}\label{meanvalue}
{\hat \mu _{kj}} = \hat \mu _{kj}^{MLE} = {({\sum\limits_{i = 1}^n {{x_{ij}}{\pi _{ik}}} }) / {{\sum\limits_{i = 1}^n {{\pi _{ik}}} }}},
\end{align}
\begin{align}\label{variance}
{\hat \sigma _{kj}} = \hat \sigma _{kj}^{MLE} = \sqrt {{{\sum\limits_{i = 1}^n {{{ [ {{\pi _{ik}} ( {{x_{ij}} - {{\hat \mu }_{kj}}}  )} ]}^2}} } / { ( {\sum\limits_{i = 1}^n {{\pi _{ik}}}  - 1} )}}}.
\end{align}

The estimation of feature weights is mainly to estimate the MI (\ref{mutualinformation}).
If $x_j$ or $x_{j'}$  is a continuous variable, the corresponding auxiliary two-valued variable is used for computing MI. $P\left( {{x_{j}}} \right)$ and $P\left( {{ x_{j'}}} \right)$ can be calculated in a similar way, and this paper only shows the calculation of $P\left( {{x_{j}}} \right)$
\begin{align}\label{pxj1}
{\hat P\left( {{x_{j}=1}} \right)} = {{\sum\limits_{i = 1}^n {{x _{ij}}} } / n}, \notag \\
 {\hat P\left( {{x_{j}=0}} \right)} = {{\sum\limits_{i = 1}^n {{{\bar x_{ij}}}} } / n},
\end{align}
where
\begin{align}\label{xbar}
{\bar x_{ij}} = \left\{ {\begin{array}{*{20}{c}}
{1{\kern 1pt} ,{\kern 1pt} {\kern 1pt} {\kern 1pt} {\kern 1pt} {\kern 1pt} {x_{ij}} = 0},\\
{0{\kern 1pt} {\kern 1pt} ,{\kern 1pt} {\kern 1pt} {\kern 1pt} {\kern 1pt} {x_{ij}} = 1}.
\end{array}} \right.
\end{align}

The estimate of $P ( {{x_j},{x_{j'}}} )$ is shown in Theorem $1$.

\textbf{Theorem 1.} For two-valued variables ${x_j}: \{ x_{1j}, x_{2j}, \ldots ,$   $  x_{nj} \}$  and  $x_{j'}: \{ x_{1j'},x_{2j'}, \ldots ,x_{nj'} \} $, $P( x_j, x_{j'} )$ can be estimated as
\begin{align}\label{pxj1xj11}
&\hat P ( {{x_j} = 1,{x_{j'}} = 1} ) = \hat P ( {{x_{j'}} = 1} ){{\hat \varphi }_n}, \notag \\
&\hat P ( {{x_j} = 0,{x_{j'}} = 1} ) = \hat P ( {{x_{j'}} = 1} ) ( {1 - {{\hat \varphi }_n}} ), \notag \\
&\hat P ( {{x_j} = 0,{x_{j'}} = 0} ) = \hat P ( {{x_{j'}} = 0} ){{\hat {\varphi '} }_n}, \notag \\
&\hat P ( {{x_j} = 1,{x_{j'}} = 0} ) = \hat P ( {{x_{j'}} = 0} ) ( {1 - {{\hat {\varphi '} }_n}} ),
\end{align}
where ${\hat \varphi _n}{\rm{ = }}\frac{{\sum\limits_{i = 1}^n {{x_{ij}}{x_{ij'}}} }}{{\sum\limits_{i = 1}^n {{x_{ij'}}} }}{\kern 1pt} $ and ${\hat \varphi '_n}{\rm{ = }}\frac{{n + \sum\limits_{i = 1}^n {{x_{ij}}{x_{ij'}}}  - \sum\limits_{i = 1}^n { ( {{x_{ij}} + {x_{ij'}}} )} }}{{n - \sum\limits_{i = 1}^n {{x_{ij'}}} }}$.

The proof of Theorem $1$ is shown in Appendix \ref{AppendixAProofofTheorem1}. Compared with MHNBM, the introduction of feature weights and the technology of clipping effectively avoids directly calculating the conditional probability ${P\left( {{{x}_{j}}\left| {{{x}_{j'}},{y} = k} \right.} \right)}$ between variables. Equation (\ref{thetakjbymle}), (\ref{meanvalue}) and (\ref{variance}) are calculations for each variable. The feature weights can be obtained through two-valued or auxiliary two-valued variables. In this paper, all the estimates of probability are double-truncated according to the above method if necessary. The above analysis is illustrated by the following Algorithm.
\begin{table}[!ht]\normalsize
\begin{center}
\label{FWMNBMAlgorithmOfflinemodeling}
\begin{tabular}{lcclc}
\toprule
 \textbf{Algorithm:} FWMNBM\\
\midrule
\textbf{Offline modeling}:\\
$Step$ 1. Divide the training data (${\pmb X,y}$) into two-valued \\
~~~~~~~~~~~ variables ${\pmb X}_t$ and continuous variables ${\pmb X}_c$.\\
$Step$ 2. Construct the auxiliary two-valued variable for \\
~~~~~~~~~~~each continuous variable according to (\ref{clippingresult}). \\
$Step$ 3. Calculate the estimates of each probability  \\
~~~~~~~~~~~via (\ref{pxj1}) and (\ref{pxj1xj11}). \\
$Step$ 4. Calculate the mutual information between \\
~~~~~~~~~~~variables and between variables and labels. \\
$Step$ 5. Calculate the weight of the feature ($F{W_{j}}$). \\
$Step$ 6. Estimate the response functions (${\hat \theta _{kj}}$) and the \\
~~~~~~~~~~~ prior probabilities (${\hat p_k}$) of two-valued variables. \\
$Step$ 7. Estimate the mean ($\hat \mu _{kj}$) and the standard \\
~~~~~~~~~~~deviation ($\hat \sigma _{kj}$) of continuous variables.\\
$Step$ 8. Build the model with the estimated parameters.\\

\textbf{Online detection}: \\
$Step$ 9. Select the sampled data and construct vector $\tilde {\pmb x}$ \\
~~~~~~~~~~~via (\ref{vecterx}).\\

$Step$ 10. Calculate ${{\pmb \varphi} _k}$, ${\phi _k}$ via (\ref{vecterthita}) and (\ref{vectermeanvar}).  \\

$Step$ 11. Calculate ${\tilde {\pmb x} \cdot {{\pmb \varphi} _k} + {\phi _k}}$ for every $k$, then \\
~~~~~~~~~~~$\mathop {\arg \max }\limits_k \left( {\tilde {\pmb x} \cdot {{\pmb \varphi} _k} + {\phi _k}} \right)$ is the predicted label. \\

 \bottomrule
\end{tabular}
\end{center}
\end{table}
\begin{figure*}[!thbp]
\centering

\subfigure[]{
\centering
\includegraphics[scale=0.52]{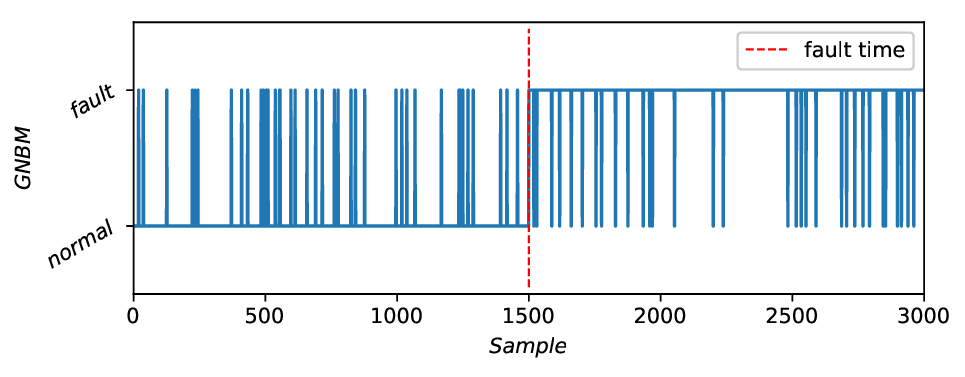}
\label{ThelabelindicatorofGNB}
}%
\subfigure[]{
\centering
\includegraphics[scale=0.52]{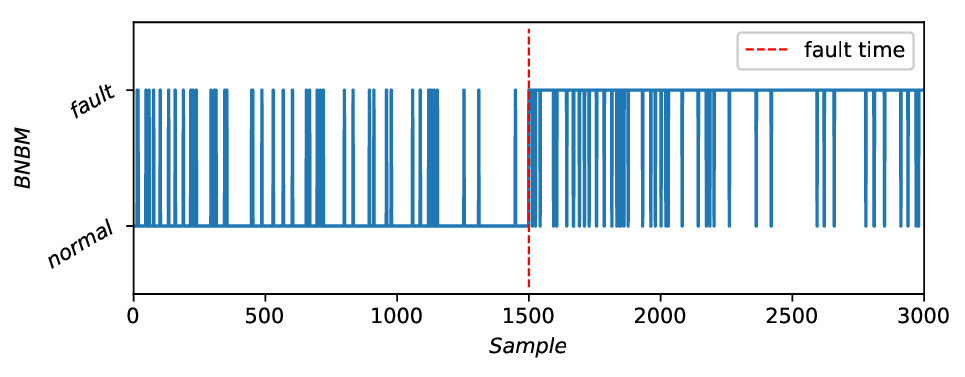}
\label{ThelabelindicatorofBNB}
}
\subfigure[]{
\centering
\includegraphics[scale=0.52]{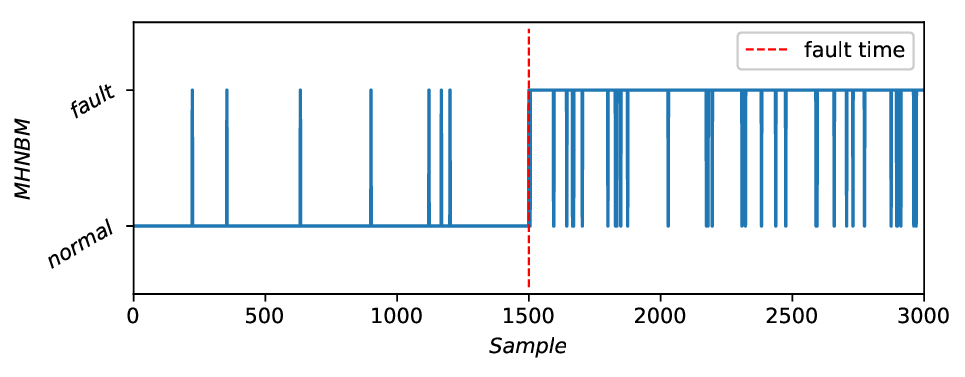}
\label{ThelabelindicatorofMHNBM1}
}%
\subfigure[]{
\centering
\includegraphics[scale=0.52]{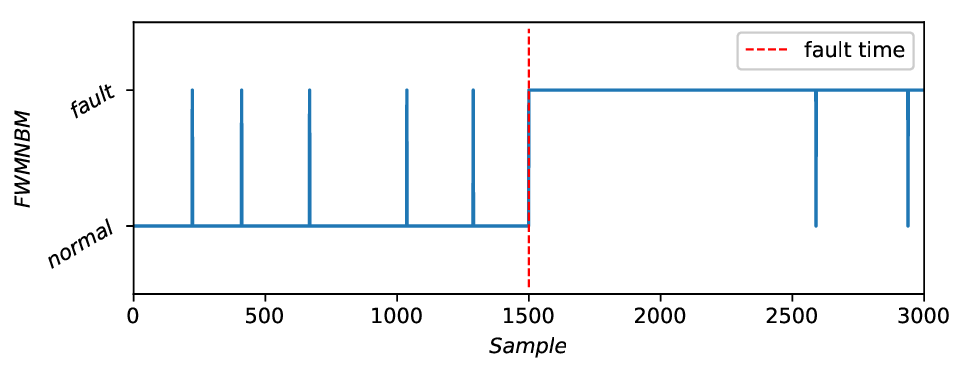}
\label{ThelabelindicatorofSUFWMNBM}
}

\centering
\caption{The label results of different methods. (a)GNBM; (b)BNBM; (c)MHNBM; (d)FWMNBM.}
\label{resultssupervisedlearning1}
\end{figure*}

\section{Simulation}
\label{Casestudies}
In this section, the numerical cases of a numerical simulation example and a practical vibration fault case of ZTPP are utilized to validate the effectiveness of FWMNBM.

\subsection{Numerical simulation}

The numerical simulation data contains 5 continuous variables and 5 two-valued variables. The means of continuous variables are shown in Table \ref{Continuousfeaturesimulationmeans} and corresponding stds are displayed in Table \ref{Continuousfeaturesimulationstds}. The two-valued variable values under different classes are depicted in Table \ref{Binaryfeaturesimulationparameterst}. In order to make the case more general, the two-valued variable values under different classes are randomly adjusted. The adjustment percentages are listed in Table \ref{Binaryfeaturesimulationparameterst}. For instance, some values of $v_6$ under normal working conditions, which are set as 0, are changed to be 1 after adjusting. Under each working condition, 1500 samples are randomly generated according to the parameters. The samples under normal 1 and fault 1 are used for training the model, and the other instances are used for testing.
\begin{table}[!htbp]
\setlength{\abovecaptionskip}{0pt}
\setlength{\belowcaptionskip}{10pt}
\centering
\caption{The preset means of continuous variables}{\label{Continuousfeaturesimulationmeans}}
\begin{tabular}{|p{1.8cm}<{\centering}|p{1.2cm}<{\centering}|p{1.2cm}<{\centering}|p{1.2cm}<{\centering}|p{1.2cm}<{\centering}|}
\hline
\multirow{2}*{\diagbox[width=1.8cm,trim=l]{variables}{means}}&\multicolumn{4}{c|}{conditions}\\ 
\cline{2-5}
&normal 1&fault 1&normal 2&fault 2\\
\hline
$v_1$&0.00&3.50&0.32&3.25\\
\hline
$v_2$&0.00&4.50&0.28&4.40\\
\hline
$v_3$&0.00&3.20&0.24&3.12\\
\hline
$v_4$&0.00&2.20&0.01&2.15\\
\hline
$v_5$&0.00&0.80&0.00&0.80\\
\hline
\end{tabular}
\end{table}
\begin{table}[!htbp]
\setlength{\abovecaptionskip}{0pt}
\setlength{\belowcaptionskip}{10pt}
\centering
\caption{The preset stds of continuous variables}{\label{Continuousfeaturesimulationstds}}
\begin{tabular}{|p{1.6cm}<{\centering}|p{1.2cm}<{\centering}|p{1.2cm}<{\centering}|p{1.2cm}<{\centering}|p{1.2cm}<{\centering}|}
\hline
\multirow{2}*{\diagbox[width=1.8cm,trim=l]{variables}{stds }}&\multicolumn{4}{c|}{conditions}\\ 
\cline{2-5}
&normal 1&fault 1&normal 2&fault 2\\
\hline
$v_1$&1.50&1.00&1.49&1.25\\
\hline
$v_2$&1.60&2.50&1.56&2.55\\
\hline
$v_3$&0.80&1.70&0.88&1.73\\
\hline
$v_4$&2.00&1.80&2.00&1.75\\
\hline
$v_5$&1.40&2.70&1.40&2.70\\
\hline
\end{tabular}
\end{table}
\begin{table}[!htb]
\setlength{\abovecaptionskip}{0pt}
\setlength{\belowcaptionskip}{10pt}
\centering
\caption{The values and  adjustment  percentage of two-valued variables}{\label{Binaryfeaturesimulationparameterst}}
\begin{tabular}{|p{2cm}<{\centering}|p{0.6cm}<{\centering}|p{1.4cm}<{\centering}|p{0.6cm}<{\centering}|p{1.4cm}<{\centering}|}
\hline
\multirow{2}*{\diagbox[width=2cm,trim=l]{variables}{conditions}}&\multicolumn{2}{c|}{normal}&\multicolumn{2}{c|}{fault}\\ 
\cline{2-5}
&values&percentage&values&percentage\\
\hline
$v_6$&0&30\%&1&30\%\\
\hline
$v_7$&1&25\%&0&25\%\\
\hline
$v_8$&0&20\%&1&20\%\\
\hline
$v_9$&0&15\%&1&15\%\\
\hline
$v_{10}$&1&10\%&0&10\%\\
\hline
\end{tabular}
\end{table}

The Gaussian naive Bayesian model (GNBM) is used for the continuous variables and the Bernoulli naive Bayesian model (BNBM) is utilized to two-valued variables. That is only $v_1,\cdots,v_5$ are used for build and test GNBM, and BNBM just utilize the information of $v_6,\cdots,v_{10}$ for modeling and verification. Different from GNBM and BNBM, MHNBM and FWHNBM are utilized for modeling and anomaly detection with both two-valued and continuous variables. The first 1500 samples of test data are normal data, and the rest are marked as faults. The test results of all above models for the testing data are depicted in Fig. \ref{ThelabelindicatorofGNB}-Fig. \ref{ThelabelindicatorofSUFWMNBM}. There are a lot of false alarms and missing faults when only continuous or two-valued variables are used, which can be seen in Fig. \ref{ThelabelindicatorofGNB} and Fig. \ref{ThelabelindicatorofBNB}. MHNBM and FWHNBM have better performance because they can simultaneously mine continuous and two-valued information at the same time. Compared to MHNBM, FWHNBM has the lower false alarm rate (FARs) and a higher fault detection rate (FDR) which are depicted in Fig. \ref{ThelabelindicatorofMHNBM} and Fig. \ref{ThelabelindicatorofFWMNBM}.

\subsection{Actual data validation}

A vibration fault of ZTPP is also used to illustrate the effectiveness of FWMNBM. At 11:35 on September 3, 2017, a hydraulic cylinder vibration fault of the primary air fan occurred, and it was recovered after 26 hours. The data, containing 495 two-valued variables and 260 continuous variables, is sampled every 5 seconds and collected from 11:35, September 1, 2017. A total of 53280 instances are collected for model training and testing.

\begin{figure*}[!htb]
	\centering
	
	\subfigure[]{
		\centering
		\includegraphics[scale=0.52]{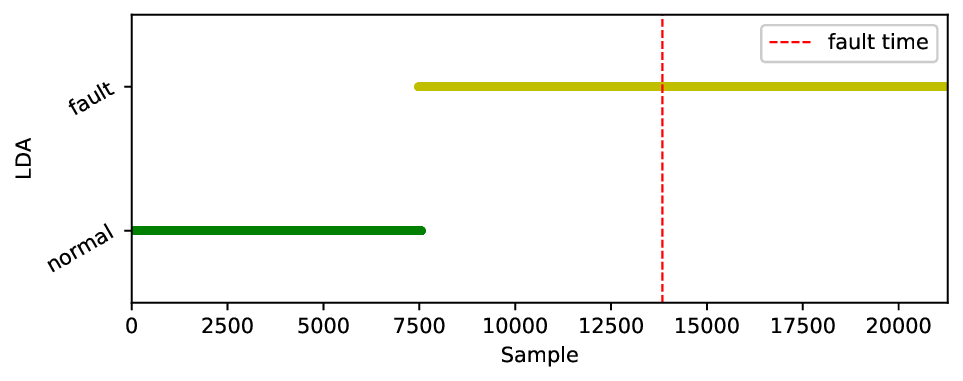}
		\label{ThelabelindicatorofLDA}
	}%
	\subfigure[]{
		\centering
		\includegraphics[scale=0.52]{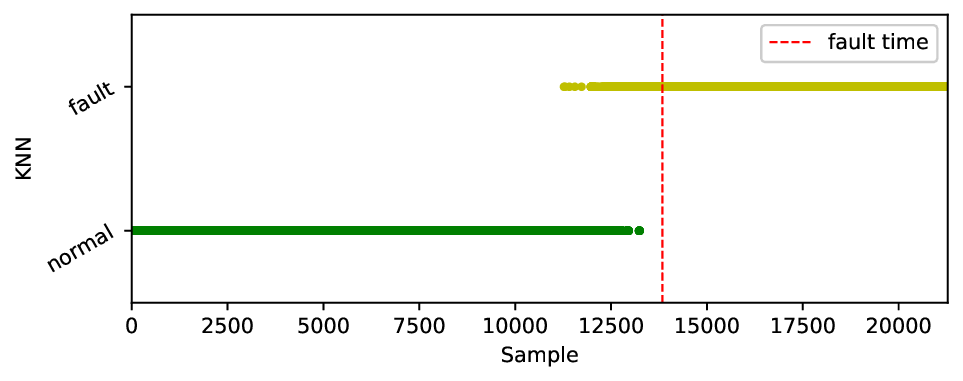}
		\label{ThelabelindicatorofKNN}
	}

	\subfigure[]{
		\centering
		\includegraphics[scale=0.52]{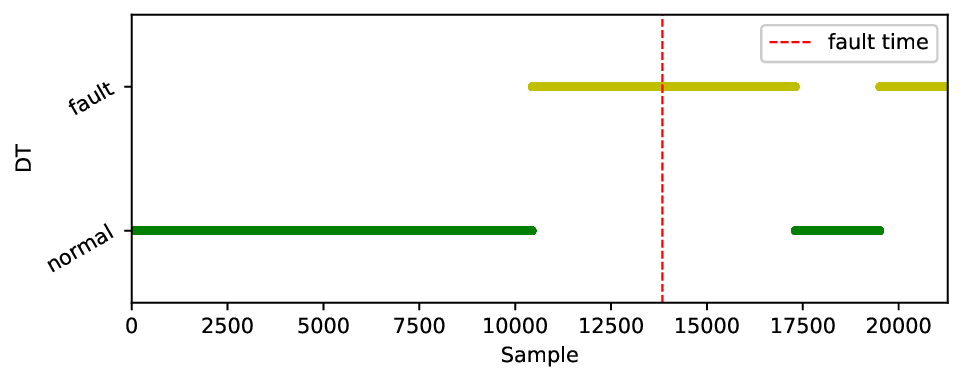}
		\label{ThelabelindicatorofRF}
	}
	\subfigure[]{
		\centering
		\includegraphics[scale=0.52]{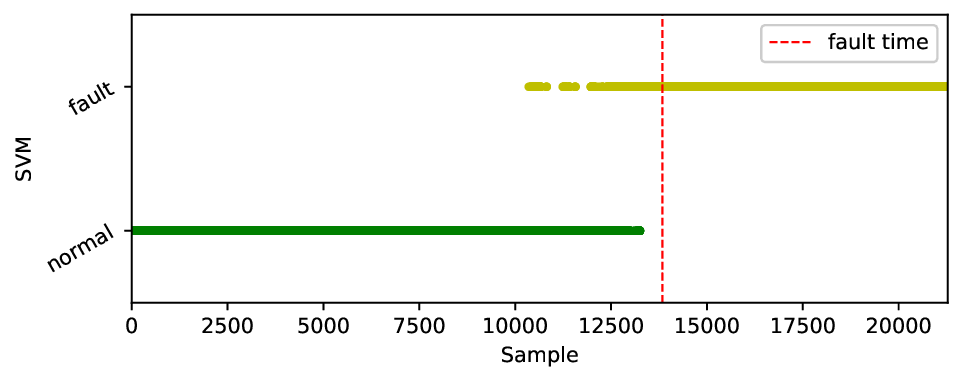}
		\label{ThelabelindicatorofSVM}
	}%

	\subfigure[]{
		\centering
		\includegraphics[scale=0.52]{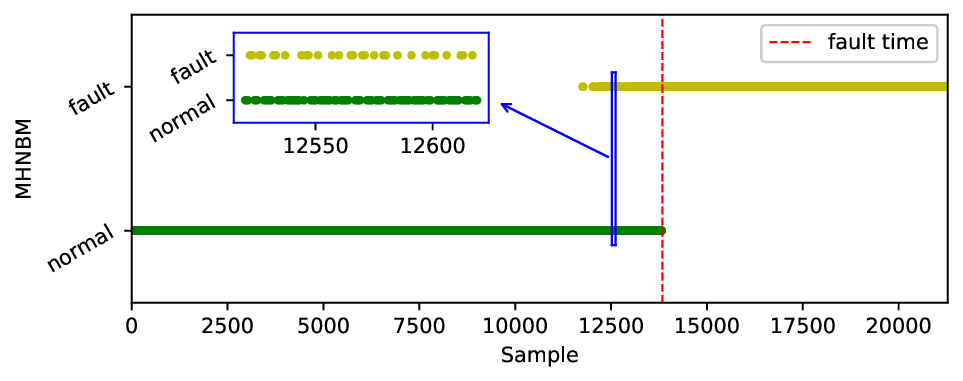}
		\label{ThelabelindicatorofMHNBM}
	}
	\subfigure[]{
		\centering
		\includegraphics[scale=0.52]{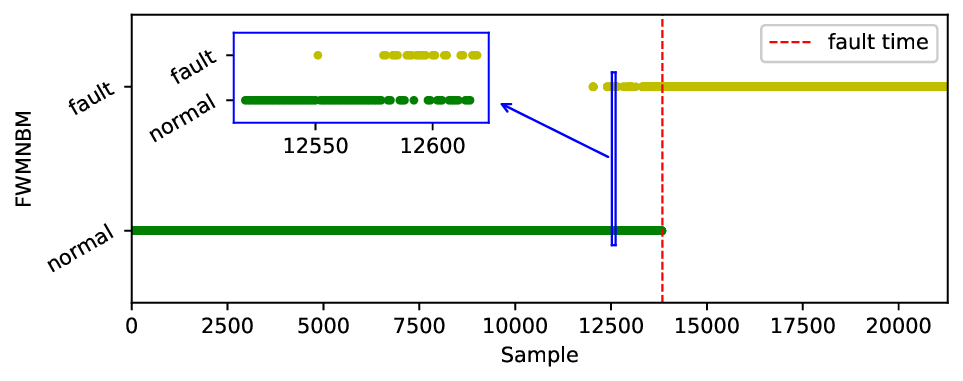}
		\label{ThelabelindicatorofFWMNBM}
	}%

	\centering
	\caption{The testing results. (a)LDA; (b)KNN; (c)DT; (d)SVM; (e)MHNBM; (f)FWMNBM.}
	\label{resultssupervisedlearning}
\end{figure*}

The first 60\% instances under normal conditions and first 60\% fault instances are utilized for modeling, and the remaining data is used for testing. In this article, we used 35 two-valued variables and continuous variables respectively. The detailed variable selection process can refer to article \cite{MinWang2020CEP}. In the traditional methods, LDA \cite{He2005A}, decision trees (DT) \cite{Sheng2004Decision}, SVM \cite{Achmad2007Support}, k-nearest neighbors (KNN) \cite{He2007Fault} are adopted to detect anomaly with the continuous variables. MHNBM and FWMNBM are used with both two-valued and continuous variables. The testing results of all methods are shown in Fig. \ref{ThelabelindicatorofLDA}-Fig. \ref{ThelabelindicatorofFWMNBM}.
\begin{table}[!htbp]
\setlength{\abovecaptionskip}{0pt}
\setlength{\belowcaptionskip}{10pt}
\setlength{\tabcolsep}{2mm}
\centering
\caption{FARs and FDRs of all methods}{\label{FDR_FAR}}
\renewcommand\arraystretch{1.2}
\begin{tabular}{|p{1.8cm}<{\centering}|p{1.8cm}<{\centering}|p{1.8cm}<{\centering}|p{1.8cm}<{\centering}|}
\hline
\textbf{Number} & \textbf{Methods} & \textbf{FARs}(\%) & \textbf{FDRs}(\%)\\
\hline
1 & LDA &45.65&100\\
\hline
2 & KNN &7.21&100\\
\hline
3 & DT &24.54&70.39\\
\hline
4 & SVM &8.82&100\\
\hline
5 & MHNBM &3.10&100\\
\hline
6 & FWMNBM &2.45&100\\
\hline
\end{tabular}
\end{table}

Excepting for DT, the performance of other methods in terms of FDR are very satisfactory. DT has omission of fault and  all methods have false alarms. In order to compare the performance of various methods, the FDRs and FARs of all methods are shown in Table \ref{FDR_FAR}. From the experimental results, the addition of two-valued variables can reduce the impact of parameter fluctuations before a fault occurs. Affected by anomaly evolution, some normal instances are misclassified into fault. However, MHNBM and FWMNBM effectively reduce FAR through combining multiple process data sources, because the advantages of both two-valued and continuous variables are taken into consideration. Among all methods, FWMNBM has the best detection performance.

\section{Conclusion}
\label{Conclusion}
 A data-driven anomaly detection method called FWMNBM is proposed with both two-valued and continuous variables. For FWMNBM, the variables more correlated to class have greater weights, which makes the more discriminating variables contribute more to the model. At the same time, FWMNBM can effectively avoid calculating the conditional probability between variables so that it can still be used when the amount of training data is not sufficient. In addition, a more effective consistent characterization method for the correlation of mixed variables is provided, and the corresponding feasibility analysis is conducted. The superior performance of FWMNBM is verified by the numerical cases of a numerical simulation example and an actual plant's case. Compared to traditional classical approaches, MHNBM and FWMNBM significantly improve the anomaly monitering performance by increasing the information of two-valued variables. Furthermore, FWMNBM has more outstanding performance because greater weights are assigned to variables with greater difference under different working conditions.

\appendices
\section{Analysis of Definition 1}
\label{AppendixAProofofDefinition1}
Definition 1 unifies the correlation analysis between variables containing both two-valued  and continuous variables by the same standard. The correlation between two-valued variables and two-valued variables or between continuous variables and continuous variables can be effectively characterized, and the original two-valued variables do not change. Therefore, the rationality of Definition 1 can be proved when a quantitative relationship exists between the correlation index of  the auxiliary two-valued variables and that of original continuous variables.

Let ${x_j} \in {\pmb X_c}$ and ${{x}_{j'}} \in {{\pmb X}_c}$, where ${{{x}'}_j}$ and ${{ x'}_{j'}}$ are the corresponding auxiliary two-valued variables. For simplicity, we assume that ${x_j}$ and ${x_{j'}}$ are standard normal distributions through standardization, that is, $E ( {{x_j}} ) = E ( {{x_{j'}}} ) = 0$, $D ( {{x_j}} ) = D ( {{x_{j'}}} ) = 1$.

The joint probability density function of ${x_j}$ and ${x_{j'}}$ is
\begin{align}\label{Thejointprobabilitydensityfunctionxjandxjj}
&f ( {{x_j},{x_{j'}}} ) \notag \\
&= \frac{1}{{2\pi \sqrt {1 - {{ [ {\rho ( {{x_j},{x_{j'}}} )} ]}^2}} }} \times \exp  \{ { - \frac{1}{{2 ( {1 - {{ [ {\rho ( {{x_j},{x_{j'}}} )} ]}^2}}  )}}} \notag \\
&  \times { [ {{{ ( {{x_j}} )}^2} + {{ ( {{x_{j'}}} )}^2} - 2\rho  ( {{x_j},{x_{j'}}}  ){x_j}{x_{j'}}} ]} \}.
\end{align}

Since
\begin{align}\label{Exjxj11}
&E ( {{{x'}_j}{{x'}_{j'}}} ) = P ( {{{x'}_j} = 1,{{x'}_{j'}} = 1} )
= P ( {{x_j} > 0,{ x_{j'}} > 0} )  \notag \\
&=\frac{1}{{2\pi \sqrt {1 - {{ [ {\rho ( {{x_j},{x_{j'}}} )} ]}^2}} }}\int\limits_0^\infty  {\int\limits_0^\infty  {\exp  \{ { - \frac{1}{{2 ( {1 - {{ [ {\rho  ( {{x_j},{x_{j'}}} )} ]}^2}} )}}} } } \notag \\
&~~~\times {[ {{{( {{x_j}})}^2} + {{( {{x_{j'}}})}^2} - 2\rho ( {{x_j},{x_{j'}}} ){x_j}{x_{j'}}}]} \}d{x_j}d{ x_{j'}} \notag \\
&=\frac{1}{{2\pi \sqrt {1 - {{[ {\rho ( {{x_j},{x_{j'}}})} ]}^2}} }}\int\limits_0^\infty  {\int\limits_0^{\frac{\pi }{2}} {r\exp \{ { - \frac{1}{{2( {1 - {{[ {\rho ( {{x_j},{x_{j'}}} )}]}^2}})}}} } } \notag \\
&~~~\times {[ {1 - \rho( {{x_j},{x_{j'}}})\sin 2\theta } ]} \}d\theta dr \notag \\
& = \frac{1}{{2\pi \sqrt {1 - {{[ {\rho ( {{x_j},{x_{j'}}})} ]}^2}} }}\int\limits_0^{\frac{\pi }{2}} {\frac{1}{{1 - \rho( {{ x_j},{ x_{j'}}})\sin 2\theta }}} d\theta \notag  \\
& = \frac{1}{{2\pi \sqrt {1 - {{[ {\rho ( {{x_j},{x_{j'}}})}]}^2}} }} \notag \\
&~~~\times \int\limits_0^{\frac{\pi }{2}} {\frac{1}{{\tan {\theta ^2} - 2\rho ( {{x_j},{x_{j'}}} )\tan \theta  + 1}}} d\tan \theta \notag \\
& = \frac{1}{{2\pi }} \{ {\frac{\pi }{2} + \arctan  [ {{{\rho  ( {{ x_j},{x_{j'}}}  )}}/{\sqrt {1 - {{ [ {\rho ( {{x_j},{x_{j'}}} )} ]}^2}} }} ]} \},
\end{align}
then
\begin{align}\label{Exjxj1}
&E ( {{{x'}_j}{{x'}_{j'}}} ) = \frac{1}{4} + \frac{1}{{2\pi }}\arcsin \rho ( {{x_j},{x_{j'}}} ) .
\end{align}

Let $P( {{{x'}_j} = 1| {{{x'}_{j'}} = 1}} ) = \varphi $.

Note that

 $P ( {{{x'}_j} = 1} ) = P ( {{x_j} > 0} ) = \frac{1}{2}$,

 $P ( {{{x'}_j} = 0} ) = P ( {{ x_j} \le 0} ) = \frac{1}{2}$,

 $E ( {{{x'}_j}} ) = 1 \times P ( {{{x'}_j} = 1} ) + 0 \times P ( {{x'_j} = 0} ) = \frac{1}{2}$,

$D ( {{{x'}_j}} ) = E{ ( {{{x'}_j} - \frac{1}{2}} )^2} = E [ {{{ ( {{{x'}_j}}  )}^2}} ] - { [ {E ( {{{x'}_j}} )} ]^2} = \frac{1}{4}$
and
$P ( {{{x'}_{j'}} = 1} ) = P ( {{{x'}_{j'}} = 0} ) = \frac{1}{2}$,

$E ( {{{x'}_{j'}}} ) = \frac{1}{2}, D ( {{{x'}_{j'}}} ) = \frac{1}{4}$.

Then we have
\begin{align}\label{meanxjxj}
&E ( {{{x'}_j}{{x'}_{j'}}} ) \vspace{1ex}
 = P ( {{{x'}_j} = 1,{{x'}_{j'}} = 1} ) \notag \\
&~~~~~~~~~~~~~= P ( {{{x'}_j} = 1} )P ( {{{x'}_j} = 1 | {{{x'}_{j'}} = 1} }  ) \vspace{1ex}
 = \frac{1}{2}\varphi, \\
&Cov ( {{{x'}_j},{{x'}_{j'}}} )
 = E [ {( {{{x'}_j} - \frac{1}{2}} ) ( {{{x'}_{j'}} - \frac{1}{2}} )} ]  \notag \\
&~~~~~~~~~~~~~~ = E( {{{ x'}_j}{{ x'}_{j'}}}) - \frac{1}{4} \vspace{1ex}
 = \frac{1}{2}\varphi  - \frac{1}{4},\\
&\rho( {{{x'}_j},{{x'}_{j'}}})
 = \frac{{Cov( {{{x'}_j},{{ x'}_{j'}}})}}{{\sqrt {D( {{{x'}_j}})D( {{{x'}_{j'}}} )} }} \notag \\
&~~~~~~~~~~~~~ = 2\varphi  - 1 \vspace{1ex} = 4E( {{{ x'}_j}{{x'}_{j'}}}) - 1 \notag \\
&~~~~~~~~~~~~~ = 4 \times[ {\frac{1}{4} + \frac{1}{{2\pi }}\arcsin \rho ( {{x_j},{ x_{j'}}})}] - 1,
\end{align}
and then
\begin{align}\label{rouxjxj}
\rho ( {{{x'}_j},{{x'}_{j'}}} ) = \frac{2}{\pi }\arcsin \rho ( {{ x_j},{x_{j'}}} ).
\end{align}

\section{Proof of Theorem $1$}
\label{AppendixAProofofTheorem1}

Let $P( {{x_j} = 1| {{x_{j'}} = 1} }) = \varphi $, $\hat P( {{x_j} = 1| {{x_{j'}} = 1} }) = {{\hat \varphi }_n}$ and $P( {{x_j} = 0| {{x_{j'}} = 0} }) = {\varphi '} $,$\hat P( {{ x_j} = 0| {{x_{j'}} = 0} } ) = {{\hat {\varphi '} }_n}$.

Then
\begin{align}\label{theConditionalprobabilitiesofallconditions}
&\hat P( {{x_j} = 0| {{x_{j'}} = 1}}) = ( {1 - {{\hat \varphi }_n}} ),  \notag \\
&\hat P( {{x_j} = 1| {{x_{j'}} = 0} } ) = ( {1 - {{\hat {\varphi '} }_n}} ).
\end{align}

Since $P( {{x_j},{x_{j'}}} )= P( {{x_{j'}}} ) P( {{x_j}|{ x_{j'}}})$. It can be learned that
\begin{align}\label{probabilitiesestimationofallconditions}
&\hat P( {{x_j} = 1,{x_{j'}} = 1}) = \hat P( {{x_{j'}} = 1} ){{\hat \varphi }_n}, \notag \\
&\hat P( {{x_j} = 0,{x_{j'}} = 1}) = \hat P( {{x_{j'}} = 1} )( {1 - {{\hat \varphi }_n}}), \notag \\
&\hat P( {{x_j} = 0,{x_{j'}} = 0} ) = \hat P( {{x_{j'}} = 0}){{\hat {\varphi '} }_n}, \notag \\
&\hat P( {{x_j} = 1,{x_{j'}} = 0} ) = \hat P( {{x_{j'}} = 0} )( {1 - {{\hat {\varphi '} }_n}} ).
\end{align}

In addition, we have \cite{2009Adaptive}
\begin{align}\label{hatvarphinjoindis}
&P( {{x_j} = {x_{1j}},{x_{j'}} = {x_{1j'}}})P( {{x_j} = {x_{2j}},{x_{j'}} = {x_{2j'}}}) \ldots  \notag \\
&P( {{x_j} = {x_{ij}},{x_{j'}} = {x_{ij'}}}) \ldots P( {{x_j} = {x_{nj}},{x_{j'}} = {x_{nj'}}}) \notag \\
& = P( {{x_{j'}} = {x_{1j'}}})P( {{x_j} = {x_{1j}}| {{x_{j'}} = {x_{1j'}}} })P( {{x_{j'}} = {x_{2j'}}})  \notag \\
&P( {{x_j} = {x_{2j}}| {{x_{j'}} = {x_{2j'}}} } ) \ldots P( {{x_{j'}} = {x_{ij'}}} ) \notag \\
&\times P( {{x_j} = {x_{ij}}| {{x_{j'}} = {x_{ij'}}} } ) \ldots P( {{x_{j'}} = {x_{nj'}}} ) \notag \\
&\times P( {{x_j} = {x_{nj}}| {{x_{j'}} = {x_{nj'}}} }) \notag \\
& = \prod\limits_{i = 1}^n {P( {{x_{j'}} = {x_{ij'}}})} \prod\limits_{i = 1}^n {P( {{x_j} = {x_{ij}}| {{x_{j'}} = {x_{ij'}}} } )}.
\end{align}

Then it can be obtained that
\begin{align}\label{pxjxjii}
&P( {{x_j} = {x_{ij}}| {{x_{j'}} = {x_{ij'}}} } ) = {\varphi ^{{x_{ij}}{x_{ij'}}}}{( {1 - \varphi })^{{x_{ij'}} - {x_{ij}}{x_{ij'}}}} \notag \\
&~~~~~~~~~~~\times {{\varphi '}^{1 + {x_{ij}}{x_{ij'}} - {x_{ij}} - {x_{ij'}}}}{( {1 - \varphi '} )^{{x_{ij}} - {x_{ij}}{x_{ij'}}}}.
\end{align}

The likelihood function is
\begin{align}\label{thelikelihoodfunctionofvarphi}
&\ell ( {\varphi ,\varphi '}) = \prod\limits_{i = 1}^n {P( {{ x_{j'}} = {x_{ij'}}} )} \prod\limits_{i = 1}^n {P( {{x_j} = {x_{ij}}| {{ x_{j'}} = {x_{ij'}}} } )} \notag \\
& = \varpi {\varphi ^{\sum\limits_{i = 1}^n {{x_{ij}}{x_{ij'}}} }}{( {1 - \varphi } )^{\sum\limits_{i = 1}^n {{x_{ij'}}}  - \sum\limits_{i = 1}^n {{x_{ij}}{x_{ij'}}} }}{{\varphi '}^{n + \sum\limits_{i = 1}^n {{x_{ij}}{x_{ij'}}} }}  \notag \\
&~~~~~~~~~~\times {{\varphi '}^{ - \sum\limits_{i = 1}^n {( {{x_{ij}} + {x_{ij'}}} )} }}{\left( {1 - \varphi '} \right)^{\sum\limits_{i = 1}^n {{x_{ij}}}  - \sum\limits_{i = 1}^n {{x_{ij}}{x_{ij'}}} }},
\end{align}
where $\varpi $ is a constant. The derivative $\frac{{\partial \ell ( \varphi ,\varphi '  )}}{{\partial \varphi }}$ of ${\ell ( \varphi,\varphi ' )}$ with respect to $\varphi $ is
\begin{align}\label{Thederivativeofvarphi}
&\frac{{\partial \ell ( \varphi, \varphi ' )}}{{\partial \varphi }} \notag \\
&=[ {\varpi {{\varphi '}^{n + \sum\limits_{i = 1}^n {{x_{ij}}{x_{ij'}}}  - \sum\limits_{i = 1}^n {( {{x_{ij}} + {x_{ij'}}} )} }}{{( {1 - \varphi '} )}^{\sum\limits_{i = 1}^n {{x_{ij}}}  - \sum\limits_{i = 1}^n {{x_{ij}}{x_{ij'}}} }}} ]\notag \\
&\times {\sum\limits_{i = 1}^n {{x_{ij}}{x_{ij'}}} }{\varphi ^{\sum\limits_{i = 1}^n {{x_{ij}}{x_{ij'}}} }}{\varphi ^{ - 1}}{\kern 1pt} {\kern 1pt} {\left( {1 - \varphi } \right)^{\sum\limits_{i = 1}^n {{x_{ij'}}}  - \sum\limits_{i = 1}^n {{x_{ij}}{x_{ij'}}} }}  \notag \\
& + [ {\varpi {{\varphi '}^{n + \sum\limits_{i = 1}^n {{x_{ij}}{x_{ij'}}}  - \sum\limits_{i = 1}^n {( {{x_{ij}} + {x_{ij'}}} )} }}{{( {1 - \varphi '} )}^{\sum\limits_{i = 1}^n {{x_{ij}}}  - \sum\limits_{i = 1}^n {{x_{ij}}{x_{ij'}}} }}} ]  \notag \\
&\times  {\varphi ^{\sum\limits_{i = 1}^n {{x_{ij}}{x_{ij'}}} }}[ {\sum\limits_{i = 1}^n {{x_{ij'}}}  - \sum\limits_{i = 1}^n {{x_{ij}}{x_{ij'}}} } ]{\kern 1pt} {\kern 1pt} {( {1 - \varphi } )^{\sum\limits_{i = 1}^n {{x_{ij'}}}  - \sum\limits_{i = 1}^n {{x_{ij}}{x_{ij'}}} }}   \notag \\
&\times  {( {1 - \varphi } )^{ - 1}}( { - 1} ).
\end{align}

Let $\frac{{\partial \ell ( \varphi,\varphi' )}}{{\partial \varphi }} = 0$, then
\begin{align}\label{thelikelihoodfunctionofvarphin}
{{\hat \varphi }_n}{\rm{ = }}\frac{{\sum\limits_{i = 1}^n {{x_{ij}}{x_{ij'}}} }}{{\sum\limits_{i = 1}^n {{x_{ij'}}} }}{\kern 1pt}.
\end{align}

The derivative $\frac{{\partial \ell ( \varphi ,\varphi ' )}}{{\partial \varphi' }}$ of ${\ell ( \varphi,\varphi ' )}$ with respect to $\varphi '$ is
\begin{align}\label{Thederivativeofvarphi1}
&\frac{{\partial \ell ( {\varphi, \varphi '} )}}{{\partial \varphi '}} = [ {\varpi {\varphi ^{\sum\limits_{i = 1}^n {{x_{ij}}{x_{ij'}}} }}{{( {1 - \varphi } )}^{\sum\limits_{i = 1}^n {{x_{ij'}}}  - \sum\limits_{i = 1}^n {{x_{ij}}{x_{ij'}}} }}} ]  \notag \\
&\times  [ {n + \sum\limits_{i = 1}^n {{x_{ij}}{x_{ij'}}}  - \sum\limits_{i = 1}^n {( {{x_{ij}} + {x_{ij'}}} )} } ]  \notag \\
&\times  {{\varphi '}^{n + \sum\limits_{i = 1}^n {{x_{ij}}{x_{ij'}}}  - \sum\limits_{i = 1}^n {( {{x_{ij}} + {x_{ij'}}} )} }}{{\varphi '}^{ - 1}}{\kern 1pt} {\kern 1pt} {( {1 - \varphi '} )^{\sum\limits_{i = 1}^n {{x_{ij}}}  - \sum\limits_{i = 1}^n {{x_{ij}}{x_{ij'}}} }}  \notag \\
& + [ {\varpi {\varphi ^{\sum\limits_{i = 1}^n {{x_{ij}}{x_{ij'}}} }}{{( {1 - \varphi })}^{\sum\limits_{i = 1}^n {{x_{ij'}}}  - \sum\limits_{i = 1}^n {{x_{ij}}{x_{ij'}}} }}} ]  \notag \\
&\times  {{\varphi '}^{n + \sum\limits_{i = 1}^n {{x_{ij}}{x_{ij'}}}  - \sum\limits_{i = 1}^n {( {{x_{ij}} + {x_{ij'}}} )} }}( {\sum\limits_{i = 1}^n {{x_{ij}}}  - \sum\limits_{i = 1}^n {{x_{ij}}{x_{ij'}}} } )  \notag \\
&\times  {( {1 - \varphi '} )^{\sum\limits_{i = 1}^n {{x_{ij}}}  - \sum\limits_{i = 1}^n {{x_{ij}}{x_{ij'}}} }}{( {1 - \varphi '} )^{ - 1}}( { - 1} ).
\end{align}

Applying $\frac{{\partial \ell \left( \varphi,\varphi'  \right)}}{{\partial \varphi' }} = 0$, it can be learned that
\begin{align}\label{thelikelihoodfunctionofvarphin1}
{{\hat \varphi '}_n} = \frac{{n + \sum\limits_{i = 1}^n {{x_{ij}}{x_{ij'}}}  - \sum\limits_{i = 1}^n {( {{x_{ij}} + {x_{ij'}}} )} }}{{n - \sum\limits_{i = 1}^n {{x_{ij'}}} }}.
\end{align}

\ifCLASSOPTIONcaptionsoff
  \newpage
\fi

\bibliographystyle{IEEEtranTIE}
\bibliography{IEEEabrv,ref_JAS_20210212}\

\end{document}